\documentclass{elsart}
\usepackage{epsfig}
    \newcommand{\ba}{\begin{eqnarray}}
    \newcommand{\ea}{\end{eqnarray}}
    \newcommand{\be}{\begin{equation}}
    \newcommand{\ee}{\end{equation}}
    \newcommand{\bp}{{\bf p}}
    \newcommand{\bx}{{\bf x}}
    \newcommand{\by}{{\bf y}}
    
    \newcommand{\tmu}{{\tilde{\mu}}}
    
    \newcommand{\hmu}{{\hat{\mu} }}
    \newcommand{\hepsilon}{{\hat{\epsilon} }}
    \newcommand{\hi}{{\hat{i} }}

    \newcommand{\psb}{\bar{\psi}}
    
    \newcommand {\bk} {{\mathbf k}}
    \newcommand {\calM} {{\mathcal M}}
    \newcommand {\calA} {{\mathcal A}}
    \newcommand {\dPhi} {{\mathcal D}\Phi}
    \newcommand {\dA} {{\mathcal D}A}
    \newcommand {\dpsi} {{\mathcal D}\psi}
    \newcommand {\dpsibar} {{\mathcal D}\bar{\psi}}
    \newcommand{\commu}[2]{\ensuremath{\left[ \; #1,\; #2\; \right]}}
    \newcommand{\aver}[1]{\ensuremath{\langle #1 \rangle}}
    \newcommand{\ket}[1]{\ensuremath{\mid #1 \;\rangle}}
    \newcommand{\bra}[1]{\ensuremath{\langle #1 \mid}}
    
    \newcommand{\matrixele}[3]{\ensuremath{\langle #1 \mid #2 \mid #3\;\rangle}}
    
    \newcommand{\AmS}{{\protect\the\textfont2
  A\kern-.1667em\lower.5ex\hbox{M}\kern-.125emS}}

\begin{document}
%\begin{CJK*}{GB}{}
\runauthor{Liu and Luo}
\begin{frontmatter}

\title{A Mean-field Calculation for the Three-Dimensional Holstein Model\thanksref{fund}}

\author[Beijing]{Chuan Liu}
\author[Beijing,State]{and Qiang Luo}
\address[Beijing]{School of Physics\\
          Peking University\\
                  Beijing, 100871, P.~R.~China}
\address[State]{State Key Laboratory for Mesoscopic Physics\\
          Peking University\\
                  Beijing, 100871, P.~R.~China}
\thanks[fund]{This work is supported by the National Natural
Science Foundation of China under Grant No. 90103006 and Grant No.
10235040 (C.L.), and the Foundation for University Key Teachers by
the Ministry of Education of China (Q.L.).}

\begin{abstract}
 A path integral representation appropriate for further Monte Carlo
 simulations is derived for the electron-phonon Holstein model in three spatial
 dimensions. The model is studied within mean-field theory.
 Charge density wave and superconducting phase transitions are discussed.
\end{abstract}
\end{frontmatter}

%\newpage

\section{Introduction}

 The $BaPb_{1-x}Bi_x O_3$ alloys \cite{Sleight} are notable
 ancestors to the high temperature superconducting
 cuprates \cite{Bednorz,Wu}. These compounds exhibit superconductivity
 in the Pb-rich composition
 range $0.05 \leq x \leq 0.3$, with a maximum $T_c \approx 13 K$
 near $x \approx 0.25$.
 A metal-semiconductor transition is observed at $x \approx 0.35$,
 with the semiconducting
 behavior continuing to the end-member compound
 $BaBiO_3$ \cite{Sleight}.

 By substitutional K doping at the A(Ba) site,
 $Ba_xK_{1-x}BiO_3$ \cite{Mattheiss1} has been found
 to exhibit superconductivity with $T_c \approx 30 K$.
 Band-structure calculations \cite{Mattheiss1,Mattheiss2} indicate
  that the conduction bands in both materials consist of $\sigma$-antibonding
  combinations of (Pb,K)-Bi($6s$) and O($2p$)
 states. These bands are gradually filled in rigid-band
 fashion with increasing $x$ until
 they are half filled in $BaBiO_3$. Here, a combination of Fermi-surface
 nesting and the strong coupling of the conduction-band states
 near $E_{F}$ to bond-stretching O displacements leads to a
 commensurate charge-density-wave (CDW) order, thereby opening a
 semiconducting gap at $E_{F}$, explaining the semiconducting
 behavior of $BaBiO_3$. The conduction bands may be well
 represented by simple
 tight-banding models. In contrast to the cuprate high temperature
 superconductors, these materials are three-dimensional and copper-free,
  rendering magnetic mechanisms inapplicable.
 Furthermore, conduction electrons couple strongly to breathing
 type oxygen phonons over a wide composition range.

 It appears to us from this picture that the three-dimensional
 Holstein Hamiltonian on a cubic lattice might constitute a
 microscopic model for these materials. The Holstein Hamiltonian
 is:
 \ba
 \label{eq:hamiltonian}
 H=-t\sum_{<\bx\by>,\sigma }
 ({c}^{\dag}_{\bx\sigma}c_{\by\sigma}+{c}^{\dag}_{\by\sigma}c_{\bx\sigma})
 +\sum_\bx [g({b}^{\dag}_\bx +b_\bx)-\mu]n_\bx
 +\omega_0 \sum_{\bx} {b}^{\dag}_\bx b_\bx \;.
 \ea
 Here, $c^{\dag}_{\bx\sigma}$ creates an electron of spin $\sigma$
 at site $\bx$, and $b^{\dag}_\bx$ is the creation operator for a
 local phonon mode of bare frequency $\omega_0$ located at site
 $\bx$. The electrons have a one-electron overlap $t$ between
 near-neighbor sites on a cubic lattice, $\mu$ is the chemical
 potential and $g$ is the
 electron-phonon coupling constant.

 For $g=0$, the free electrons in the three-dimensional Holstein
 model have a tight binding band structure:
 $\epsilon(\bk)=-2t(\cos{k_1} +\cos {k_2} +\cos {k_3})$.
 It has perfect nesting only at half-filling. As noted above, this
 band structure describes $BaPb_{1-x}Bi_x O_3$ and $Ba_xK_{1-x}BiO_3$
 quite well. The Einstein phonon in the Holstein model corresponds to
 the oxygen breathing mode phonon in real materials. When x increases,the
  rigid-band fashion filling
 of $BaPb_{1-x}Bi_x O_3$ and $Ba_xK_{1-x}BiO_3$
 corresponds to the variation of the chemical potential $\mu$ in
 the Holstein model which determines band filling. All these
 features indicate the connections between the Holstein model
 and these real materials. It should be
 noted that the boson in the Holstein model could also represent an
 "exciton", a "plasmon", or any other intermediate bosons. The
 Holstein model therefore has a wider range of applicabilities.

 The idea that a single Hamiltonian~(\ref{eq:hamiltonian})
 describes both the
 semiconducting and superconducting phases of $BaPb_{1-x}Bi_x O_3$
 and $Ba_xK_{1-x}BiO_3$ is very appealing. The basic electron-electron
 interaction arising from the exchange of phonons is given by
 \be
 \label{eq:retard}
 V(i\omega_m)=-\frac{2g^2\omega_0}{\omega^2_m+\omega^2_0}
 \ee
 with $\omega_m=2m\pi T$. In the limit $\omega_0/t\gg 1$ with
 $g^2/\omega_0$
 finite, the Holstein model goes over to a negative-U Hubbard
 model with an effective
 $U$ equal to $-2g^2/\omega_0$ \cite{Scalettar}.
 Whereas the basic electron-electron interaction in
 a Hubbard model is instantaneous and local, as is characterized by
 a constant $U$, that in the Holstein model
 is retarded, i.e., $V(i\omega_m)$
 depends on frequency (see Eq.~(\ref{eq:retard})).
 This retardation effect will play an important role
 in subsequent discussions.

 While reports on investigations for the three-dimensional Holstein
 model have been rare, the two-dimensional Holstein model has been
 previously investigated by a number of techniques,
 including quantum Monte Carlo simulations \cite{Scalettar}. Some of
 the conclusions from these studies are expected to hold in three dimensions.
 According to these studies, two competing types of  correlations
 exist in the Holstein model, $s$-wave superconducting and
 Peierls CDW. At and near half-filling, CDW long range order prevails, further
 away from half-filling, $s$-wave superconductivity dominates.
 As $\omega_0/t$ increases, the boundary separating CDW and
 superconducting phases
 moves toward half-filling. In the negative $U$ Hubbard model,
 which is a limiting case of the Holstein model, CDW can only be
 stabilized at half-filling. It is interesting to note that as
   $\omega_0/t$ is reduced and retardation becomes important, a
 region of filling around the half-filled band is dominated by CDW
 correlations. It is reasonable to assume that a CDW phase is
 insulating. We thus expect that, unlike the Hubbard model,
 an insulating (semiconducting)
 phase can be stabilized away from half-filling due to
 retardation in the Holstein model.

 As pointed out by Mattheiss and Hamann \cite{Mattheiss2}, the
 mechanism by which the semiconducting properties of
 $BaPb_{1-x}Bi_x O_3$ and $Ba_xK_{1-x}BiO_3$ are extended over
 intermediate range of compositions is less well understood. The
 semiconducting behavior at half-filling, i.e., $BaBiO_3$, may be
 understood from a combination of Fermi-surface nesting and strong
 coupling to breathing mode oxygen phonons. Both ingredients
 are necessary to open a semiconductor gap at the Fermi surface.
 The gap should disappear quickly with the loss of Fermi surface
 nesting away from hall-filling. Weber \cite{Weber} proposed that
 this semiconducting behavior was due to a combination of static
 incommensurate ``breathing-type'' charge density waves and chemical
 ``ordering waves'' on the Pb-Bi sublattice. This mechanism, however,
 has some difficulties with band-structure calculations and
 perturbative arguments \cite{Mattheiss2}.

 It thus appears from these discussions that retardation in the
 Holstein model might be responsible for the stabilization of the
 semiconducting behavior away from half-filling. $Ba_xK_{1-x}BiO_3$
 simply has a phonon frequency higher than that of $BaPb_{1-x}Bi_x O_3$.
 Consequently, the metallic regime of these alloys is extended closer to
 half-filling, where the deformation potentials for breathing-type
 O displacements and the electron-phonon interaction are at
 maximum, thus resulting in a higher $T_c$. As superconductivity
 occurs near a CDW transition, the coherence length is expected to
 be short, in agreement with experimental observations.

 Several previous studies \cite{Bickers,Zhang} have used
 certain types of negative-$U$ Hubbard models to describe
 $BaPb_{1-x}Bi_x O_3$ systems. The CDWs in Hubbard and Holstein
 models differ in one respect. CDW in the former is complete in the sense
 that alternating lattice sites are doubly occupied
 and empty; while in the latter, charge is only partly
 transferred with an amount set by retardation. Infrared
 \cite{Hair}
 and x-ray-photoemission-spectroscopy \cite{Wertheim} measurements suggest only
 marginal differences in the charge distributions at the Bi
 sites \cite{Mattheiss2}, lending further support to the Holstein model
 description of these materials.

 Symmetry arguments \cite{Scalettar1} can be used to classify electronic phase
 transitions in the three-dimensional Holstein model. At
 half-filling, superconducting and CDW channels are degenerate. In
 this case the two-component superconducting order parameter
 (complex scalar) and the one-component CDW order parameter (real
 scalar) may be considered jointly as a three-component vector
 order parameter. Away from half-filling the degeneracy between the
 superconducting and CDW phases is broken. Thus phase transitions
 of CDW at half-filling, away from half-filling, and
 superconductivity belong to the universality classes of the
 three-dimensional Heisenberg, Ising, and XY models respectively.
 The corresponding phase transitions in  $BaPb_{1-x}Bi_x O_3$ and
 $Ba_xK_{1-x}BiO_3$ are thus identified. It will be interesting to
 explore consequences of this classification in these systems.

 \section{Path integral representation of the model}

 To obtain a path
 integral representation for the (grand) canonical partition function
 $\Xi=Tr\exp[-\beta H]$, one first expresses the bosonic creation
 and annihilation operators
 in terms of a bosonic field $\Phi_\bx$ and the corresponding
 conjugate momenta $\Pi_\bx$ defined as:
 \be
 \label{eq:phi-pi}
 b_\bx=\sqrt{{\omega_0\over 2}}
 \left(\Phi_\bx+i{\Pi_\bx\over\omega_0}\right) \;\;,
 \;\;
 b^{\dagger}_\bx=\sqrt{{\omega_0\over 2}}
 \left(\Phi_\bx-i{\Pi_\bx\over\omega_0}\right) \;\;.
 \ee
 It is easy to see that $\Phi_\bx$ and  $\Pi_\bx$ defined
 above satisfy the usual commutation relations:
 $\commu{\Phi_\bx}{\Pi_\by}=i\delta_{\bx\by}$.
 The Hamiltonian of the model now becomes:
 \ba
 H&=&-t\sum_{\bx,\mu,\sigma}
 {c}^{\dag}_{\bx\sigma}
 (c_{\bx+\hat{\mu}\sigma}+c_{\bx-\hat{\mu}\sigma})
 -\sum_\bx [g\sqrt{2\omega_0}\Phi_\bx+\mu]n_\bx
 \nonumber \\
 &+&\sum_\bx {\Pi^2_\bx\over 2}+{\omega^2_0\over 2}\Phi^2_\bx \;.
 \ea
 The inverse temperature
 $\beta\equiv 1/T$ is then divided into $N_t$ slices in
 the ``temporal'' direction, each of size $a=\beta/N_t$.
 Inserting the completeness relation
 for the fermionic coherent states and the bosonic variables
 at any given time slice labeled by $\tau$:
 \ba
 \label{eq:both-complete}
 &&\int\prod_{\bx}\left[d\Phi_{\tau,\bx}\prod_{\sigma}
 \left(d\bar{\psi}_{\tau,\bx\sigma}d\psi_{\tau,\bx\sigma}
 \right)\right]
 \exp\left(-\sum_{\bx\sigma}
 \bar{\psi}_{\tau+1,\bx\sigma} \psi_{\tau,\bx\sigma}\right)
 \nonumber \\
 &\times&\ket{\Phi_{\tau,\bx};\psi_{\tau,\bx\sigma}}
 \bra{\bar{\psi}_{\tau+1,\bx\sigma};\Phi_{\tau,\bx}}=1\;\;.
 \ea
 the partition function can be transformed into:
 \[
 \Xi=\int\dPhi\dpsibar\dpsi\prod^{N_t-1}_{\tau=0}
 \matrixele{\bar{\psi}_\tau;\Phi_\tau}{e^{-aH}}{\Phi_{\tau+1};\psi_\tau}
 e^{-\sum_{\bx\sigma}\bar{\psi}_{\tau+1,\bx\sigma}\psi_{\tau,\bx\sigma}}
 \;,
 \]
 where the bosonic and fermionic integrals are over all
 field components:
 $\Phi_{\tau,\bx}$, $\psi_{\tau,\bx\sigma}$ and
 $\bar{\psi}_{\tau,\bx\sigma}$ with $\tau=0,1,\cdots,N_t-1$.
 In the temporal direction, periodic and anti-periodic boundary
 conditions are understood for
 the field $\Phi$ and $\bar{\psi}$, respectively.

 The matrix element in the above expression can be evaluated
 easily using the properties of the coherent states and the
 fact that the temporal lattice spacing $a\sim 0$. It is
 convenient to introduce a re-scaled field $A_x$ via:
 \be
 A_x=\hat{g}\sqrt{2\omega_0}\Phi_x \;\;,
 \ee
 where we have used  the notation $x$ to label
 the four-dimensional site $(\tau,\bx)$, and
 $\hat{g}$ is the dimensionless coupling measured
 by $t$, i.e.\ $\hat{g}\equiv g/t$. Similarly, we will
 use the notation: $\hat{\mu}=\mu/t$ and $\hat{\omega}_0=\omega_0/t$.
 Finally, the partition function of the
 model is given by \cite{liu00,zhu-liu01}:
 \ba
 \Xi&=&\int\dA
 \exp\left(- \sum_{x,y}A_x
 {(-\hat{\partial}^2_0+\omega^2_0a^2)_{xy} \over 4\hat{g}^2\omega_0 a}A_{y} \right)
 \det({\mathcal M}[A]^2)
 \;\;, \nonumber \\
 {\mathcal M}[A]_{xy} &=&
 at\sum^{3}_{i=1}
 (\delta_{x+\hi,y}+\delta_{x-\hi,y})
 +[1+at(\hat{\mu}+A_{x})]\delta_{xy}
 -\zeta_{x_0}\delta_{x-\hat{0},y}
 \;\;,
 \ea
 where we have used $x=(\bx,s)$ to collectively label $3+1$ dimensional
 lattice sites. The sign function
 $\zeta_{x_0}=\pm$ is present to insure the anti-periodic
 boundary condition for the fermion fields in the temporal
 direction.

 It is noted that the real fermion matrix ${\mathcal M}$ is {\em not}
 Hermitean. However, since the matrix ${\mathcal M}$ is real
 and the up and down spin electrons share the same fermion matrix,
 the effective action is positive definite
 for {\em any} value of
 the chemical potential. Therefore, this path integral
 representation for the Holstein model
 is suitable for Monte Carlo simulations \cite{liu00,zhu-liu01}
 without the infamous ``sign problem''.

 \section{The charge density wave phase}

 To perform a mean-field analysis for the possible charge density wave
 long range order, we assume that the effective action:
 \be
 \label{eq:seff}
 S_{eff}[A_x]=\sum_{x,y}A_x
 {(-\hat{\partial}^2_0+\omega^2_0a^2)_{xy} \over 4\hat{g}^2\omega_0 a}A_{y}
 -2Tr\log \calM[A_x] \;\;,
 \ee
 has a minimum at the ``ferrimagnetic'' background field
 $A^{(MF)}_x=\hat{M}+(-)^\bx\hat{\Delta}$, characterized by
 the magnetization parameter $\hat{M}$ and the staggered
 magnetization parameter $\hat{\Delta}$. In
 this background field, the fermion matrix $\calM$, which
 will be denoted as $\calM^{(0)}$, has a simple block diagonal form
 in four-momentum space:
 \be
 \calM_{pq}[A^{(MF)}]\equiv\calM^{(0)}_{pq}=
 [1-e^{-ip_0}+at(\tmu-\hat{\epsilon}_\bp)]\delta_{pq}
 +at\hat{\Delta}\delta_{q,p+Q} \;\;,
 \ee
 where $p$ and $q$ are two
 four-momenta, the temporal components of which being
 the usual fermionic Matsubara frequencies.
 The parameters $\tmu$ and $\hat{\epsilon}_\bp$ designate $\hat{\mu}+\hat{M}$
 and $\epsilon_\bp/t$, respectively,
 and $Q$ equals $(0,\pi,\pi,\pi)$.
 As we will  show shortly, a non-vanishing staggered magnetization
 order parameter $\hat{\Delta}$ signals
 a CDW phase in the system while the value
 of $\hat{M}$ represents the ``renormalization effect''
 of the chemical potential.

 We decompose the scalar field $A_x$ as:
 \[
 A_x=A^{(MF)}_x+\calA_x=\hat{M}+(-)^\bx\hat{\Delta}+\calA_x \;\;,
 \]
 and expand the effective action~(\ref{eq:seff}) around
 the mean-field solution $A^{(MF)}_x$. The fact that
 $A^{(MF)}_x$ being a saddle point of the effective action
 insures that the term linear in $\calA_x$ is absent in
 this expansion. Therefore, to second order in $\calA_x$, we
 obtain:
 \ba
 S_{eff}[A]&=& S^{(0)}_{eff}[A^{(MF)}]+S^{(2)}_{eff}[\calA]\;\;,
 \nonumber \\
 S^{(0)}_{eff}[A^{(MF)}]&=&
 {\Omega \omega_0a(\hat{M}^2+\hat{\Delta}^2)
 \over 4\hat{g}^2}
 -2Tr\log \calM[A^{(MF)}] \;\;,
 \nonumber \\
  S^{(2)}_{eff}[\calA_x]&=&
  Tr\left(\calM^{(0)-1}\delta\calM\calM^{(0)-1}\delta\calM\right)
  \;\;.
 \ea
 This serves as our starting point for the mean-field calculation.
 In this paper, only the lowest order mean-field results
 will be studied. These results are contained in $S^{(0)}_{eff}$.

 Since the mean-field fermion matrix $\calM^{(0)}$ only couples
 momentum $p$ with momentum $p+Q$, the lowest order effective
 action $S^{(0)}\equiv-\log\Xi^{(0)}$ can be easily obtained:
 \ba
 \label{eq:s0}
 S^{(0)}_{eff}[A^{(MF)}]&=&
 {\Omega \omega_0a(\hat{M}^2+\hat{\Delta}^2) \over 4\hat{g}^2}
 \nonumber \\
 &-&\sum_p\log\left[\left(1+at\tmu-e^{-ip_0}\right)^2
 -a^2t^2\hat{E}_\bp\right] \;\;,
 \ea
 where we have introduced the energy spectrum in the CDW phase:
 \be
 \label{eq:spectrum}
 \hat{E}^2_\bp=\hat{\epsilon}^2_\bp+\hat{\Delta}^2 \;\;.
 \ee
 We therefore see that, if a non-zero solution for the order
 parameter $\hat{\Delta}$ exists, the energy spectrum develops
 an energy gap characterized by $\hat{\Delta}$.

 It turns out that the summation over the temporal component
 $p_0$ in eq.~(\ref{eq:s0}) can be performed analytically.
% using techniques summarized in appendix~\ref{sec:sum-p0}.
 Therefore, we obtain:
 \ba
 S^{(0)}_{eff}&=&
 {\Omega \omega_0a(\hat{M}^2+\hat{\Delta}^2) \over 4\hat{g}^2}
 \nonumber \\
 &-&\sum_\bp\log\left[1+\left(1+at(\tmu-\hat{E}_\bp)\right)^{N_t}\right]
 \left[1+\left(1+at(\tmu+\hat{E}_\bp)\right)^{N_t}\right] \;\;.
 \ea
 In most cases, one is only concerned with the so-called
 continuum limit: $a\rightarrow 0$ with $N_ta=\beta$ being fixed.
 Then,
 \[
 \left(1+at(\tmu+\hat{E}_\bp)\right)^{N_t}
 \sim \exp\left(\beta t(\tmu+\hat{E}_\bp)\right)\;\;.
 \]
 The free energy per lattice site: $f\equiv -\log\Xi/(\beta V_3)$,
 apart from irrelevant constant terms, is thus obtained
 \footnote{We use $V_3$ to designate the number of sites of the
 three dimensional lattice.}
 to the lowest order in mean-field theory in
 the continuum limit:
 \be
 f^{(0)}={\omega_0\over 4\hat{g}^2}(\hat{M}^2+\hat{\Delta}^2)
 -{1\over V_3\beta}\sum_{\bp}
 \log[\cosh(\beta t\tmu) +\cosh(\beta t\hat{E}_\bp)] \;\;.
 \ee

 The chemical potential $\mu$ is related to the total
 electron number via: $\aver{N}=\partial\log\Xi /(\beta\partial\mu)$.
 Using the doping fraction $x=\aver{N}/V_3-1$ and taking the
 continuum limit, this equation reads:
 \be
 \label{eq:mu}
 x={1\over V_3}\sum_\bp
 {\sinh(\beta t\tmu)  \over
 \cosh(\beta t\tmu) +\cosh(\beta t\hat{E}_\bp)} \;\;.
 \ee
 It is seen that the solution $\tmu$ will have
 the same sign as $x$, in particular, at
 half-filling ($x=0$), the solution to Eq.~(\ref{eq:mu})
  is: $\tmu=0$.

 The requirement that the background field $A^{(MF)}_x$ being a minimum
 of the effective action can be utilized to derive the
 so-called gap equations. The equation for $\hat{M}$ yields the
 solution:
 \be
 \label{eq:solution-M}
  \hat{M}=2(1+x)\hat{g}^2/\hat{\omega}_0 \;\;,
 \ee
 where $1+x$ is given by eq.~(\ref{eq:mu})
 ,while the equation for $\hat{\Delta}^2$ in the continuum limit reads:
 \be
 \label{eq:delta}
 {\hat{\omega}_0\over 2\hat{g}^2} ={1\over V_3}\sum_\bp
 {\sinh(\beta t\hat{E}_\bp) /\hat{E}_\bp \over
 \cosh(\beta t\tmu) +\cosh(\beta t\hat{E}_\bp)} \;\;,
 \ee
  At a given doping level $x$, one needs to use both Eq.~(\ref{eq:mu})
 and Eq.~(\ref{eq:delta}) to solve
 for $\tmu$ and $\hat{\Delta}$.
 The critical temperature
 $T_c=1/\beta_c$ for the CDW phase is obtained from
 Eq.~(\ref{eq:delta})  and Eq.~(\ref{eq:mu}) by setting the
 order parameter $\hat{\Delta}$ to zero, or equivalently, by setting
 $E_\bp=\epsilon_\bp$ for a given value of $x$:
 \be
 \label{eq:Tc}
  {\hat{\omega}_0\over 2\hat{g}^2} ={1\over V_3}\sum_\bp
 {\sinh(\beta_c t\hat{\epsilon}_\bp) /\hat{\epsilon}_\bp \over
 \cosh(\beta_c t\tmu) +\cosh(\beta_c t\hat{\epsilon}_\bp)} \;\;.
 \ee

 It is also instructive to calculate the ensemble average of
 the local electron number operator
 $n_\bx\equiv \sum_\sigma c^{\dag}_{\bx\sigma}c_{\bx\sigma}$.
 To lowest order in mean-field theory, it is given by:
 \[
 \aver{n_\bx}=2+
 2\aver{\calM^{(0)-1}_{\tau_0,\bx;\tau_0+1,\bx}}
 \;\;.
 \]
 Keeping $a$ finite and after some algebra, one obtains:
 \be
 \aver{n_\bx}={\aver{N}\over V_3}
 +(-)^\bx{\hat{\omega}_0\hat{\Delta} \over 2 \hat{g}^2}
 + O(a)
 ={\hat{\omega}_0 \over 2 \hat{g}^2}
 \left[\hat{M}+(-)^\bx\hat{\Delta}\right]+ O(a)\;\;,
 \ee
 As we claimed before, a non-vanishing order parameter
 $\hat{\Delta}$ opens up an energy gap in the spectrum
 and signals a charge density wave order in
 the system.

 Another useful quantity to
 characterize the CDW phase transition is the CDW
 susceptibility $\chi_{CDW}$ defined by:
 \be
 \beta\chi_{CDW}={a\over V_3}\sum_{\bx,\by,\tau}
 (-)^{\bx-\by}\aver{n_{\tau,\bx}n_{0,\by}}
 \;\;.
 \ee
 By virtue of translational invariance, one obtains:
 \[
 \chi_{CDW}={1\over N_t\Omega}\sum_{x,y}
 e^{iQ(x-y)}\aver{\psb\psi(x)\psb\psi(y)}
 \;\;.
 \]
 Since the fermion propagator for up and down spins are
 identical, we have:
 \be
 \chi_{CDW}=-{2\over N_t\Omega}\sum_{x,y}
 e^{iQ(x-y)}\aver{\psi(x)\psb(u)\psi(y)\psb(x)}
 \;\;.
 \ee
 This expression can be represented in momentum space by
 a loop diagram. We will calculate the susceptibility
 in the phase where $\Delta=0$ for simplicity. To lowest order in mean-field
 theory, the susceptibility is given by:
 \be
 \chi^{(0)}_{CDW}=-{2\over N_t\Omega}\sum_{p}
 {1\over \left[1+at(\tmu-\hat{\epsilon}_\bp)-e^{-ip_0}\right]
 \left[1+at(\tmu+\hat{\epsilon}_\bp)-e^{-ip_0}\right]}
 \;\;,
 \ee
 where we have used the fact that:
 $\hat{\epsilon}_\bp=-\hat{\epsilon}_{\bp+Q}$.
 Using similar techniques as before, the summation over
 $p_0$ can be performed analytically, and
 in the continuum limit, the above equation is reduced to:
 \be
 \chi^{(0)}_{CDW}={1\over V_3}\sum_\bp
 \left({1\over\beta\epsilon_\bp}\right)
 {\sinh(\beta\epsilon_\bp) \over \cosh(\beta t\tmu)
 +\cosh(\beta\epsilon_\bp)} \;\;.
 \ee

 A better approximation, the
 Random Phase Approximation (RPA),
 for the CDW susceptibility can
 be obtained by summing over a geometric series of $\chi^{(0)}_C$
 with one phonon exchange. In this approximation, the
 CDW susceptibility becomes:
 \be
 \label{eq:rpa}
 \chi^{(RPA)}_{CDW} ={\chi^{(0)}_{CDW} \over
 1-(2\beta t\hat{g}^2/\hat{\omega_0})\chi^{(0)}_{CDW}}
 \;\;,
 \ee
 which becomes divergent exactly at the critical temperature
 $T_c$, as is evident by comparing with Eq.~(\ref{eq:Tc}).

 \section{The superconducting phase transition}

 In the previous section, we have discussed the possibility of
 the CDW phase transition in the Holstein model. This phase transition
 is characterized  by
 a long range order parameter $\Delta$ which is diagonal
 in electron spins. It turns out that the system also
 permits a phase with long range order which is
 off-diagonal in the space of the electron spins.
 This phase is the
 conventional BCS-type superconducting phase with
 Cooper pairing. To see such a phase transition,
 the phonon field is integrated out from the path
 integral representation of the partition function.
 Then, another set of boson fields, namely
 a real scalar field $\Phi_x$,
 a complex scalar field $\Delta_x$ and its complex
 conjugate $\Delta^*_x$, are integrated in.
 Introducing the four component fermion field notation:
 \be
 \Psi_x=\left(
 \begin{array}{c}
 \psi_{1x} \\\ \psi_{2x} \\ \bar{\psi}_{1x} \\ \bar{\psi}_{2x}
 \end{array}\right)\;\;,
 \ee
 the action of the model can be written as:
 \ba
 S&=&{1\over 2}\Psi_{\alpha x}
 {\mathcal D}[\Delta,\Phi]_{\alpha x;\beta y} \Psi_{\beta y}
 \nonumber \\
 &+&{\omega_0 a\over 2\hat{g}^2}\sum_x |\Delta_x|^2
 +\Phi_x\left({-\hat{\partial}^2_0+\omega^2_0a^2 \over
 4\hat{g}^2\omega_0a}\right)_{xy}\Phi_y \;\;.
 \ea
 where the index $\alpha$ and $\beta$ run from $1$ to $4$ and
 the anti-symmetric fermion matrix
  ${\mathcal D}[\Delta,\Phi]_{\alpha x;\beta y}$ is given by:
 \be
 \left(
 \begin{array}{cc|cc}
 0 & -at\Delta^{*}_x\delta_{xy} &
 -{\mathcal M}^{(0) T}_{xy}
 -{\hat{\partial}_0\over\hat{\omega}_0}\Phi_x\delta_{xy} & 0 \\
 \mbox{\ } & 0 & 0 &
 -{\mathcal M}^{(0) T}_{xy}
 -{\hat{\partial}_0\over\hat{\omega}_0}\Phi_x\delta_{xy} \\
 \hline
 \mbox{\ } & \mbox{\ } & 0 & at\Delta_x\delta_{xy} \\
 \mbox{\ } & \mbox{\ } & \mbox{\ } & 0
 \end{array}\right) \;\;,
 \ee
 where the matrix ${\mathcal M}^{(0)}_{xy}$ is given by:
 \be
 {\mathcal M}^{(0)}_{xy} =
 at\sum^{3}_{i=1}
 (\delta_{x+\hi,y}+\delta_{x-\hi,y})
 +(1+at\hat{\mu})\delta_{xy}
 -\zeta_{x_0}\delta_{x-\hat{0},y}
 \;\;.
 \ee
 The effective action of the system is obtained by integrating
 out the fermion fields, resulting in the Pfaffian (the square
 root of the determinant) of the corresponding fermion
 matrix ${\mathcal D}[\Delta,\Phi]$:
 \be
 S_{eff}={\omega_0 a\over 2\hat{g}^2}\sum_x |\Delta_x|^2
 +\Phi_x\left({-\hat{\partial}^2_0+\omega^2_0a^2 \over
 4\hat{g}^2\omega_0a}\right)_{xy}\Phi_y
 -{1\over 2}Tr\log{\mathcal D}[\Delta,\Phi] \;.
 \ee

 We now proceed to search for a constant mean-field solution
 of the type: $\Delta_x=\hat{\Delta}_{BCS}$ and $\Phi_x=\Phi_0$
 which minimizes the effective action. In this case, the
 fermion matrix ${\mathcal D}$ is diagonal in four-momentum
 space, i.e. ${\mathcal D}_{pq}={\mathcal D}^{(0)}(p)\delta_{pq}$
 and it is given by:
 \be
 {\mathcal D}^{(0)}(p)=\left(
 \begin{array}{cc|cc}
 0 & -at\hat{\Delta}^{*}_{BCS} & -{\mathcal M}^{(0)}(p)^* & 0 \\
  at\hat{\Delta}^{*}_{BCS} & 0 & 0 & -{\mathcal M}^{(0)}(p)^* \\
 \hline
 {\mathcal M}^{(0)}(p) & 0 & 0 & at\hat{\Delta}_{BCS} \\
 0 & {\mathcal M}^{(0)}(p) & at\hat{\Delta}_{BCS} & 0
 \end{array}\right) \;\;,
 \ee
 with ${\mathcal M}^{(0)}(p)=1+at(\hmu-\hepsilon_\bp)-e^{-ip_0}$.
 Therefore, to lowest order in mean-field theory, the
 effective action is found to be:
 \ba
 {1\over\Omega}S^{(0)}_{eff}&=&{\omega_0 a\over 2\hat{g}^2} |\hat{\Delta}_{BCS}|^2
 +\left({\omega_0a \over
 4\hat{g}^2}\right)\Phi^2_0 \nonumber \\
 &-&{1\over \Omega}\sum_p \log\left[
 |1-e^{-ip_0}+at(\hmu-\hepsilon_\bp)|^2
 + (at)^2|\hat{\Delta}_{BCS}|^2
 \right]\;\;.
 \ea
 Since the fermion contribution is independent of $\Phi_0$, it is
 immediately seen that the solution for $\Phi_0$ is $\Phi_0=0$.
 The summation
 over Matsubara frequencies can be performed analytically.
 In the continuum limit the result is:
 \be
 {1\over\Omega}S^{(0)}_{eff}=
 {\omega_0 a\over 2\hat{g}^2} |\hat{\Delta}_{BCS}|^2
 -{\beta\mu\over N_t}
 -{1\over \Omega}\sum_\bp\log\left(
 2+e^{-\beta E_\bp}+e^{\beta E_\bp}
 \right) \;\;,
 \ee
 where the spectrum $E_\bp$ is given by:
 \be
 E_\bp=\sqrt{(\mu-\epsilon_\bp)^2+t^2\hat{\Delta}^2_{BCS}}
 \;\;.
 \ee
 Chemical potential $\mu$ is determined from band filling.
 Using the doping fraction parameter $x$, the chemical
 potential satisfies the following equation:
 \be
 x={1\over V_3}\sum_\bp
 \left(\mu-\epsilon_\bp \over E_\bp\right)
 \tanh\left({\beta E_\bp\over 2}\right)\;\;.
 \ee
 It is noted that, at half-filling $x=0$ and the solution
 is $\mu=0$ due to the symmetry of $\epsilon_\bp$.
 The superconducting energy gap $\hat{\Delta}_{BCS}$
 is obtained from the famous BCS gap equation:
 \be
 {\hat{\omega} \over\hat{g}^2}={1\over V_3}\sum_\bp
 \left({t\over E_\bp}\right)
 \tanh\left({\beta E_\bp\over 2}\right)
 \;\;.
 \ee
 The critical temperature for the superconducting phase
 transition is obtained by setting the BCS gap parameter
 $\hat{\Delta}_{BCS}$ to zero in the above equation:
 \be
 \label{eq:Tc_BCS}
 {\hat{\omega}_0 \over\hat{g}^2}={1\over V_3}\sum_\bp
 \left({t\over |\mu-\epsilon_\bp|}\right)
 \tanh\left({\beta_c |\mu-\epsilon_\bp|\over 2}\right)
 \;\;.
 \ee
 The fermion propagator is given by the inverse matrix
 of ${\mathcal D}$ which again is diagonal in
 momentum space. If we denote:
 \be
 \aver{\Psi_{\alpha x}\Psi_{\beta y}}=
 {1\over \Omega}\sum_p e^{ip(x-y)}
 [{\mathcal D}^{(0)}]^{-1}_{\alpha\beta}(p) \;\;,
 \ee
 then the matrix $[{\mathcal D}^{(0)}]^{-1}(p)$ is given by:
 \be
 [{\mathcal D}^{(0)}]^{-1}(p)={1\over D(p)}
  \left(
 \begin{array}{cc|cc}
 0 & at\hat{\Delta}_{BCS} & {\mathcal M}^{(0)}(p)^* & 0 \\
 -at\hat{\Delta}_{BCS} & 0 & 0 & {\mathcal M}^{(0)}(p)^* \\
 \hline
 -{\mathcal M}^{(0)}(p) & 0 & 0 & -at\hat{\Delta}^*_{BCS} \\
 0 & -{\mathcal M}^{(0)}(p) & at\hat{\Delta}^*_{BCS} & 0
 \end{array}\right) \;\;,
 \ee
 with $D(p)=|{\mathcal M}^{(0)}(p)|^2+|at\hat{\Delta}_{BCS}|^2$.

 We can also calculate the superconducting susceptibility
 defined by:
 \be
 \chi_{BCS}={1\over N_t\Omega}\sum_{x,y}
 \aver{\Psi_{1x}\Psi_{2x}\Psi_{4y}\Psi_{3y}}
 \;\;.
 \ee
 For simplicity, we will evaluate the superconducting
 susceptibility in the phase with vanishing
 $\hat{\Delta}_{BCS}$ (high temperature phase).
 After some algebra, we obtain to the lowest order:
 \be
 \chi^{(0)}_{BCS}={1\over V_3}\sum_\bp
 {1\over \beta|\mu-\epsilon_\bp|}
 \tanh\left({\beta|\mu-\epsilon_\bp|\over 2}\right)
 \;\;.
 \ee
 Under RPA, the susceptibility becomes:
 \be
 \label{eq:rpa_BCS}
 \chi^{(RPA)}_{BCS} ={\chi^{(0)}_{BCS} \over
 1-(\beta t\hat{g}^2/\hat{\omega}_0)\chi^{(0)}_{BCS}}
 \;\;,
 \ee
 which becomes divergent exactly at $\beta_c$ determined by
 Eq.~(\ref{eq:Tc_BCS}).

 \section{Results and discussions}

 \begin{figure}[htb]
 \begin{center}
 \includegraphics[width=12.0cm]{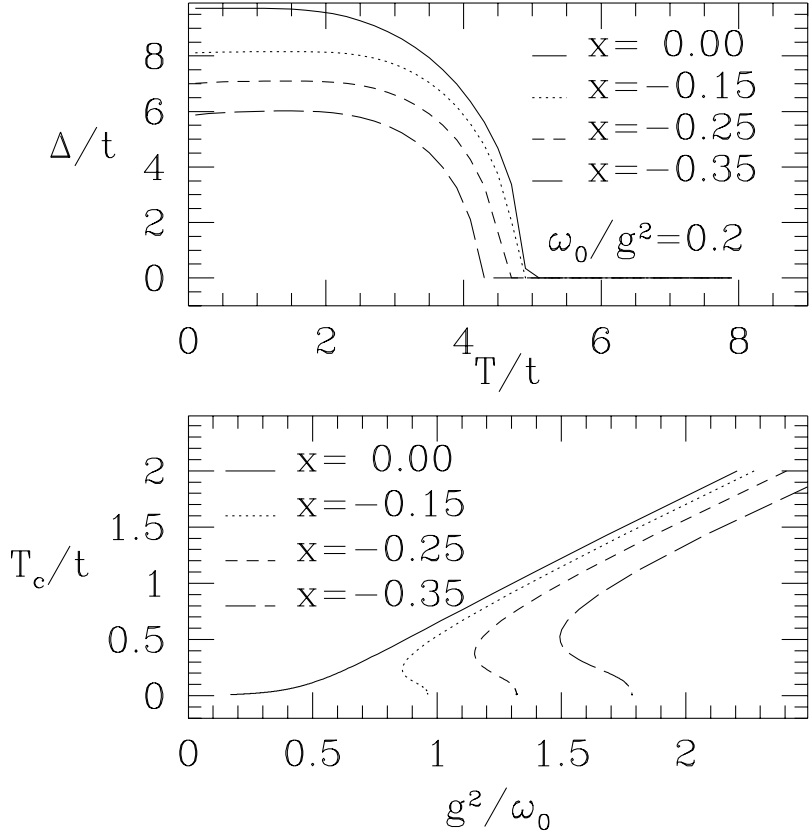}
 \caption{In the lower half of the figure,
 the critical temperatures for the CDW phase
 transition are plotted as a function
 of $g^2/\omega_0$ for four different doping levels:
 $x=0.00,-0.15,-0.25,-0.35$.  In the upper half of the figure,
 the order parameters for the CDW phase transition
 are plotted as a function
 of the temperature $T/t$ for a given value of
 $\omega_0/g^2=0.2$ for four different doping levels:
 $x=0.0,-0.15,-0.25,-0.35$. \label{fig:Tc_delta1}}
 \end{center}
 \end{figure}

\begin{figure}[htb]
 \begin{center}
 \includegraphics[width=12.0cm]{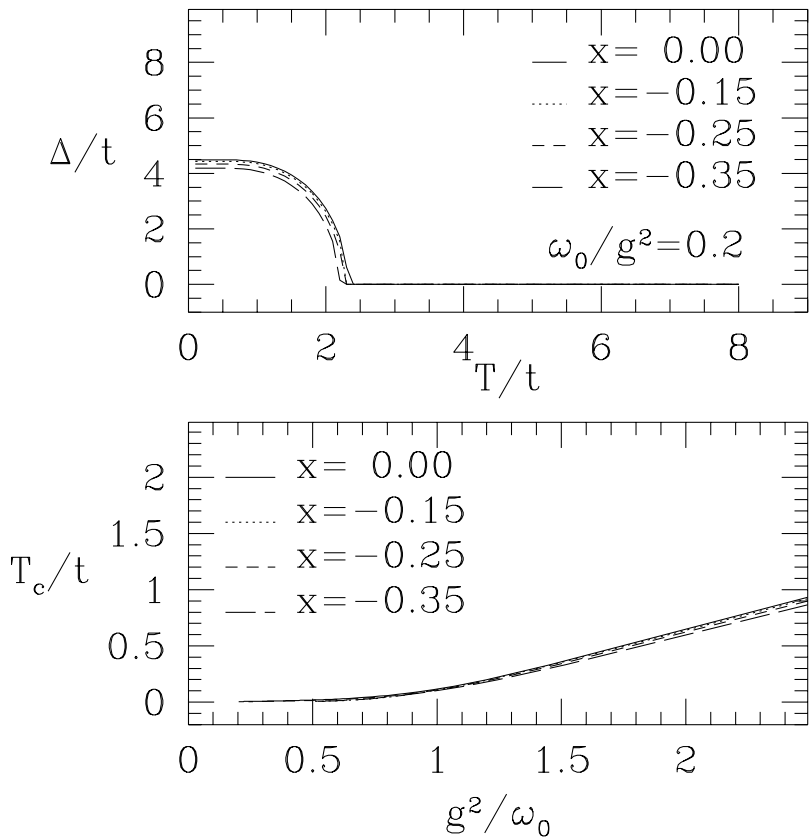}
 \caption{In the lower half of the figure,
 the critical temperatures for the BCS superconducting phase
 transition are plotted as a function
 of $g^2/\omega_0$ for four different doping levels:
 $x=0.00,-0.15,-0.25,-0.35$.  In the upper half of the figure,
 the order parameters for the BCS superconducting phase transition
 are plotted as a function
 of the temperature $T/t$ for a given value of
 $\omega_0/g^2=0.2$ for four different doping levels:
 $x=0.0,-0.15,-0.25,-0.35$. \label{fig:Tc_delta2}}
 \end{center}
 \end{figure}

 Mean-field equations discussed in previous sections can be
 evaluated numerically and the results for the critical
 temperatures
 and the order parameters are summarized
 in Fig.~\ref{fig:Tc_delta1} and Fig.~\ref{fig:Tc_delta2}.
 In the lower half of the figures,
 we plot the critical temperatures
 as a function of $\hat{g}^2/\hat{\omega}_0$ for four
 different doping levels: $x=0.00,-0.15,-0.25,-0.35$.
 The critical temperatures for the superconducting phase transition are
 represented by the four lines that increase with
 $\hat{g}^2/\hat{\omega}_0$ near the bottom of Fig.~\ref{fig:Tc_delta2}.
 They show rather weak dependence on doping. The corresponding
 lines for the CDW phase transition illustrate a much stronger dependence
 on doping compared to the superconducting critical temperatures.
 It is noted that, for a given value of
 $\hat{g}^2/\hat{\omega}_0$, the
 critical temperatures for the CDW phase transition
 decrease as the doping is increased.
 At exactly half filling, the critical temperature for
 the CDW phase transition is always higher than that of
 the superconducting phase transition. While the issue of the stability
 of the CDW and superconducting phases can only be addressed by a more
 sophisticated calculation, here we simply take the magnitude of
 the critical temperatures and order parameters
 as a measure of the tendency that the system is likely to be in one of these
 phases. Thus, we expect the system to be in a CDW phase at half filling. Away from
 half filling, the critical temperature for the
 CDW phase transition eventually
 becomes smaller than that of the superconducting phase
 transition. This is a signal that the CDW phase will eventually
 becomes unstable when the system is doped away from
 half filling. The superconducting phase will become
 dominant at low temperatures.

 In the upper half of the figures,
 we plot the order parameters
 $\hat{\Delta}=\Delta/t$ for the CDW
 and the superconducting phase transitions as a function of
 temperature $T/t$ at
 $\hat{\omega}_0/\hat{g}^2=0.2$ for the same
 four different doping levels.
 It is seen that, for a given
 value of $\hat{\omega}_0/\hat{g}^2$, both
 order parameters decrease as the doping is increased.
 However, the order parameter for the CDW phase decreases
 much more rapidly than the superconducting order parameter
 when the system moves away from half filling.
 These features indicate that the CDW phase is most
 favored at half filling, in agreement with our qualitative
 expectations discussed in the introduction.

 \section{Conclusions}

 In this paper, we present our mean-field analysis of the
 three dimensional Holstein model. A path integral representation
 of the model is obtained which is suitable for further Monte
 Carlo simulations. The critical temperatures and the order
 parameters are obtained within mean-field approximation for both
 the CDW phase and BCS superconducting phase transitions. Our results are
 consistent with the picture that the CDW phase is most favored at half-filling. Away
 from half filling, the usual BCS superconducting phase becomes
 important. Higher order calculations are necessary to further
 explore the competition between different electronic phase
 transitions.

%\end{CJK*}

\begin{thebibliography}{99}
 \bibitem{Sleight} A.W. Sleight, J.L. Gillson, and P.E. Bierstedt,
 Solid State Commun. {\bf17}, 27 (1975).
 \bibitem{Bednorz} J.G. Bednorz and K.A. Muller, Z. Phys.
 {\bf B 64}, 189 (1986); J.G. Bednorz, M. Takashigi, and K.A.
 Muller, Europhys. Lett {\bf 3}, 379 (1987).
 \bibitem{Wu} M.K. Wu, J.R. Ashburn, C.J. Torng,
 P.H. Hor, R.L. Meng, L. Gao, Z.J. Huang, Y.Q. Wang, and C.W. Chu,
 Phys. Rev. Lett. {\bf 58}, 908 (1987).
 \bibitem{Mattheiss1} L.F. Mattheiss, E.M. Gyorgy, and D.W. Johnson, Jr.
 Phys. Rev. {B 37}, 3745 (1988).
 \bibitem{Mattheiss2} L.F. Mattheiss, and D.R. Hamann, Phys. Rev. {B 28}
 4227 (1983); L.F. Mattheiss, and D.R. Hamann, Phys. Rev. Lett.
 {\bf 60}, 2681 (1988).
 \bibitem{Scalettar} R.T. Scalettar, N.E. Bickers, and D.J.
 Scalapino, Phys. Rev. {B 40}, 197 (1989).
 \bibitem{Weber} W. Weber, Jpn. J. Appl. Phys. {\bf 26}, Suppl. 3,
 981 (1987)
 \bibitem{Bickers} N.E. Bickers, R.T. Scalettar, and D.J. Scalapino,
  Int. J. Mod. Phys. {\bf B 1}, 687 (1987)
\bibitem{Zhang} Zhang Li-yuan, Solid State Commu. {\bf 70}, 1065
(1989)
\bibitem{Scalettar1} R.T. Scalettar, E.Y. Loh, J.E. Gubernatis, A. Moreo,
S.R. White, D.J. Scalapino, R.L. Sugar, and E. Dagotto, Phys. Rev.
Lett. {\bf 62}, 1407 (1989).
\bibitem{Hair} J.Th.W. de Hair and G. Blasse, Solid State Commun.
{\bf 12}, 727 (1973)
\bibitem{Wertheim} G.K. Wertheim, J.P. Remeika, and D.N.E.
Buchanan, Phys. Rev. {B 26}, 2120 (1982)
%
 \bibitem{liu00} C.~Liu, Chin.~Phys.~Lett. {\bf 17}, 574 (2000)
%
 \bibitem{zhu-liu01} X.~Zhu and C.~Liu, Commun.~Theor.~Phys., {\bf
 36}, 625 (2001)
\end{thebibliography}
\end{document}